\documentclass[journal,10pt]{IEEEtran}
\usepackage{mathptmx}
\usepackage{balance}
\usepackage{ulem}
\usepackage[numbers,sort&compress]{natbib}
\usepackage{amsthm,amsmath,amssymb,thmtools}
\usepackage{mathrsfs}
\usepackage{graphicx}
\usepackage{color}
\usepackage{enumitem}
\makeatletter
\newcommand*\bigcdot{\mathpalette\bigcdot@{.5}}
\newcommand*\bigcdot@[2]{\mathbin{\vcenter{\hbox{\scalebox{#2}{$\m@th#1\bullet$}}}}}
\makeatother
\usepackage{epstopdf}
\usepackage{booktabs}
\usepackage{bm}
\normalsize
\usepackage{caption}
\usepackage{subcaption}
\declaretheoremstyle[
  headfont=\normalfont\itshape,
  headpunct=\textup{:},
  bodyfont=\normalfont,
  headindent=1em
]{myremark}
\usepackage[algo2e,ruled,linesnumbered]{algorithm2e}
\makeatletter
\newcommand{\nosemic}{\renewcommand{\@endalgocfline}{\relax}}
\newcommand{\dosemic}{\renewcommand{\@endalgocfline}{\algocf@endline}}
\let\oldnl\nl
\newcommand{\nonl}{\renewcommand{\nl}{\let\nl\oldnl}}
\makeatother
\SetAlCapHSkip{0pt}
\SetKwIF{If}{ElseIf}{Else}{if}{then}{else~if}{else}{end~if}
\usepackage{float}
\usepackage{setspace}

\newlength{\textfloatsepsave} 

\usepackage{algpseudocode}

\begin{document}
\title{ LLM-Empowered IoT for 6G Networks:  Architecture, Challenges, and Solutions}
\author{Xiaopei~Chen,~\IEEEmembership{Student Member,~IEEE}, Wen~Wu,~\IEEEmembership{Senior Member,~IEEE},  Liang~Li,~\IEEEmembership{Member,~IEEE}, Fei~Ji,~\IEEEmembership{Member,~IEEE} 
	
	\thanks{\scriptsize{\textit{Corresponding author: Wen Wu and Fei Ji.}}}
	\thanks{Xiaopei Chen is with the School of Future Technology, South China University of Technology, Guangzhou 511442, China, and also with the Frontier Research Center, Pengcheng Laboratory, Shenzhen 518000, China (e-mail: ftchenxp@mail.scut.edu.cn). }

	\thanks{Wen Wu and Liang Li are with the Frontier Research Center, Pengcheng Laboratory, Shenzhen 518000, China (e-mail: \{wuw02, lil03\}@pcl.ac.cn). }
	\thanks{Fei Ji is with the School of Electronic and Information Engineering, South China University of Technology, Guangzhou 510640, China (e-mail: eefeiji@scut.edu.cn). }
}
\markboth{ }
{}

\maketitle

\begin{abstract} The Internet of Things (IoT) in the sixth generation (6G) network is envisioned to evolve towards intelligence, ubiquity, and self-optimization. Large language models (LLMs) have demonstrated remarkable generalization capabilities across diverse domains, including natural language processing, computer vision, etc. In this article, we propose an LLM-empowered IoT architecture for 6G networks to achieve intelligent autonomy and facilitate diversified IoT applications.  On one hand,  LLM-based solutions are tailored to satisfy the quality of service requirements of  IoT applications, i.e., LLM for 6G IoT. On the other hand,  LLMs are expected to be deployed in IoT environments through edge fine-tuning methods and edge inference techniques to support IoT applications, i.e.,  LLM on 6G IoT. Furthermore, we propose a memory-efficient split federated learning framework for LLM fine-tuning on heterogeneous IoT devices to alleviate memory consumption on both IoT devices and the edge server. Finally, a case study is presented, followed by a discussion about open issues.
\end{abstract}

\begin{IEEEkeywords}
	Internet of Things, large language model, split federated learning.
\end{IEEEkeywords}

\IEEEpeerreviewmaketitle

\section{Introduction}
Compared to the fifth-generation (5G), sixth-generation (6G) networks are expected to bring revolutionary breakthroughs in ultra-low latency, ultra-high bandwidth, intelligence, and enhanced autonomy  \cite{ITU2023}. 6G networks will rely on cutting-edge technologies such as terahertz communication, intelligent reflective surfaces, space-air-ground integrated networks, and edge intelligence to realize the new era of the interconnected intelligence of all things \cite{9651548, 9749222}. These features will not only promote the upgrading of human-computer interaction and data-driven applications but also profoundly change the computing mode and intelligence level of the Internet of Things (IoT). As a core application of the 6G network,  IoT will usher in a profound transformation in three aspects: massive device access, intelligent data processing, and automated decision-making. 

Riding the wave of advancements in artificial intelligence (AI), we are witnessing the eruption of transformer-based large language models (LLMs) such as GPT and LLaMA, which have demonstrated remarkable generalization capabilities across diverse domains, including natural language processing, computer vision, etc. With the advent of the 6G era, LLMs are expected to enhance the intelligence of IoT. 

The success of LLMs is fundamentally driven by their scalability. LLMs demonstrate remarkable task-level scalability by adopting a pretraining–fine-tuning paradigm. In the pretraining phase, the model is trained on massive and diverse corpora to acquire general language understanding and reasoning capabilities. This is followed by task- or domain-specific fine-tuning, where only a small amount of labeled data is needed to adapt the model to downstream tasks. This paradigm enables LLMs to be flexibly extended to support a wide range of IoT applications without redesigning model architectures.  Unlike earlier approaches, LLMs exhibit continuous improvements in accuracy and generalization as data volume or parameter count increases, all while maintaining the same underlying simple algorithms and architectures. The scaling law formalizes this phenomenon, demonstrating how the performance of transformer-based models improves in a predictable manner as model size and dataset scale expand \cite{kaplan2020scaling}. The IoT can provide massive data to motivate the development of modern LLMs.

The integration of LLMs into 6G IoT presents a compelling architectural vision. With the ability to perform language understanding, semantic reasoning, and task generalization, LLMs can serve as a powerful cognitive engine to elevate traditional IoT systems toward more human-centric and autonomous applications in the 6G era. Unfortunately, most existing LLM products are heavily dependent on cloud computing, which comes with significant drawbacks, including excessive latency, high bandwidth costs, and serious privacy concerns. On the other hand, the enormous memory requirements and computational demands of LLMs make direct deployment on resource-constrained IoT devices infeasible.

In this article, we propose an LLM-empowered IoT architecture for 6G networks to achieve intelligent autonomy while supporting advanced IoT applications. LLMs are pushed to the edge of the 6G network to achieve the synergy between LLMs and IoT. The synergy of LLM and IoT in the proposed architecture is two-fold: On one hand, LLMs are applied to empower IoT applications and enhance IoT network management, namely LLM for 6G IoT. On the other hand, LLMs are expected to be deployed to support IoT applications in IoT environments through edge fine-tuning methods and edge inference techniques, namely LLM on 6G IoT. Specifically, parameter-efficient fine-tuning reduces the computational resource requirements for LLM fine-tuning, and distributed learning and collaborative inference can be adapted to the limited memory and computational power of IoT devices. Furthermore, we propose an SFL framework for memory-efficient fine-tuning over heterogeneous IoT devices, where the server maintains a full LLM and the corresponding LoRA modules are selectively fine-tuned in a sequential manner for each IoT device. The case study demonstrates that the proposed scheme can reduce memory footprint and training time while achieving comparable performance.

 \begin{figure*}[h]
 	\centering
 	\includegraphics*[width=0.8\textwidth]{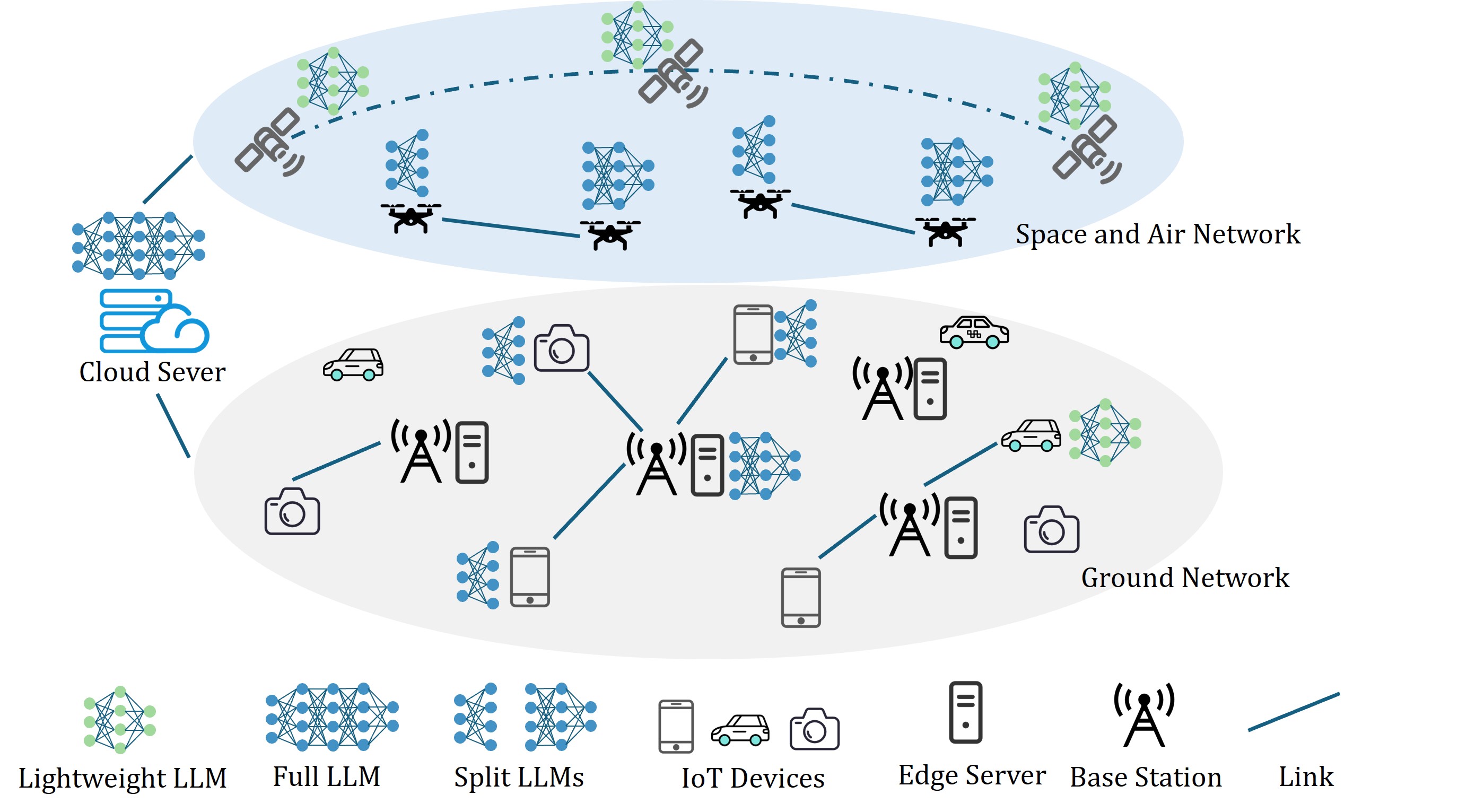}\\
 	\caption{LLM-empowered IoT architecture in 6G networks.}\label{fig.architecture}
 \end{figure*}

The remainder of this paper is organized as follows. Section II discusses some expected applications of 6G IoT, and then the LLM-empowered IoT architecture is proposed. The basic ideas of LLM for 6G IoT and LLM on 6G IoT are presented in Sections III and IV, respectively. Section V presents the memory-efficient SFL framework. Section VI provides a case study. Open issues are discussed in Section VII, and the conclusions are summarized in Section VIII.

\section{LLM-Empowered IoT in 6G Networks}
\subsection{6G IoT Applications}
The advent of 6G technology promises unprecedented transformation in IoT applications, pushing the boundaries of connectivity, intelligence, and responsiveness beyond current capabilities. Leveraging ultra-high data rates, sub-millisecond latency, seamless global coverage, and integrated sensing capabilities, 6G IoT will empower a variety of advanced use cases across multiple sectors. 
\begin{itemize} 
	\item  \textbf{Smart Transportation}:  Enhanced vehicular communications (V2X) will enable real-time interactions between vehicles, infrastructure, pedestrians, and cloud systems, improving traffic management and fully autonomous driving.
	\item  \textbf{Industrial IoT}:   Ultra-reliable low-latency communications will revolutionize industrial automation, enabling precise robotic control, remote operations, and instant feedback, thus boosting manufacturing efficiency and resilience. 	
	\item  \textbf{Smart City}: Cities will integrate comprehensive sensing networks with cameras, environmental sensors, and smart infrastructures, enabling proactive urban management. Real-time analytics will lead to enhanced public safety, efficient resource use, and improved quality of life.
	\item  \textbf{eHealth and Human-Machine Fusion}: Healthcare will advance dramatically through remote robotic surgery, wearable and implantable health monitoring devices, and sophisticated brain-computer interfaces, supported by 6G's ultra-low latency and high throughput.
	\item  \textbf{Integrated Space-Air-Ground-Sea IoT}: 6G will integrate terrestrial networks, aerial platforms, satellites, and maritime sensors, ensuring global connectivity, robust communication in extreme environments, and improved disaster management.
\end{itemize}

\begin{figure*}[h]
	\centering
	\includegraphics*[width=0.9\textwidth]{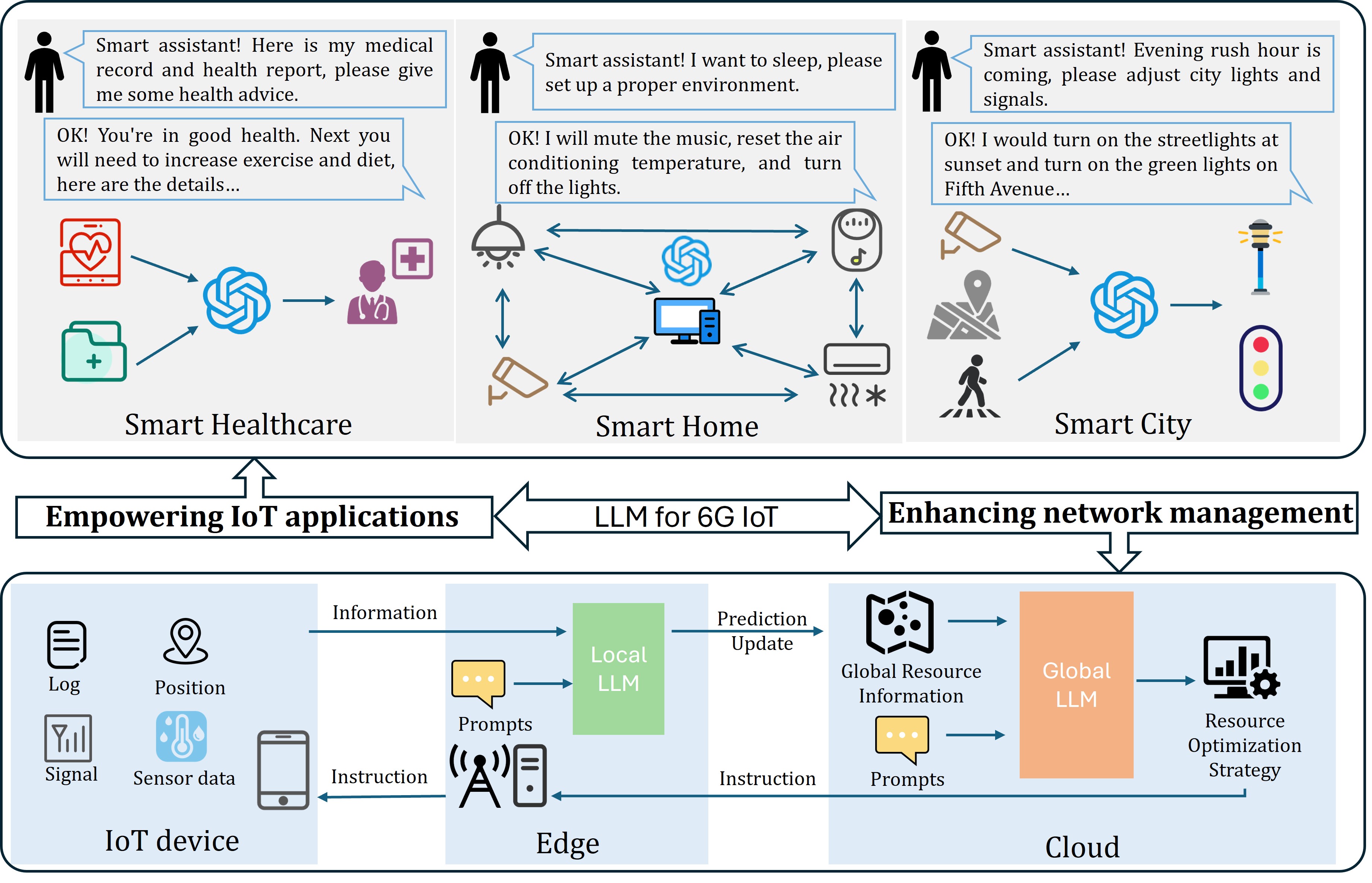}\\
	\caption{LLM for 6G IoT.}\label{fig.architecture}
\end{figure*}

\subsection{Large Language Model} 
The development of LLMs has undergone rapid evolution over the past decade. Early advances were marked by the introduction of the Transformer architecture in 2017, which replaced recurrent networks with self-attention mechanisms and enabled more efficient modeling of long-range dependencies. In 2018, with the outstanding performance of BERT and GPT-1 in NLP tasks, the pre-training-fine-tuning paradigm became mainstream.  Since 2019, the advent of LLMs (such as GPT-2 and GPT-3 from OpenAI, BERT variants and T5 from Google, and XLNet from CMU/Google) has led to major advances in text generation, zero-shot generalization, and in-context reasoning.  These developments were further accelerated by models such as Google's PaLM, DeepMind's Chinchilla and Gopher, Meta's LLaMA series, and Anthropic's Claude, where each contributes to the scaling and refinement of LLM capabilities. 

An LLM usually consists of an embedding layer, multiple transformer layers, and an output layer \cite{chang2024survey}. The impressive adaptability of LLMs stems from the pre-training and fine-tuning paradigm. The pre-training phase uses massive unsupervised data (publicly available data) to learn generalized linguistic knowledge through self-supervision. The pre-training phase requires a significant amount of resources, which include not only the computing processors like GPUs and TPUs but also the memory, energy, and network bandwidth. For example, the pre-training of LLaMa-2-70B takes two trillions of data tokens, 1.7 million GPU hours, and consumes $2.5 \times 10^{12}$ J of energy. Subsequently, the pre-trained models are fine-tuned to acquire domain-specific knowledge, thus enhancing their generalization capabilities and applicability to specific tasks. Compared to pre-training, the data used for fine-tuning is a small amount of task-specific data (license required), and the resources consumed are less.

A pressing challenge has arisen alongside the rapid advancement of LLMs: projections indicate that high-quality public datasets may be exhausted by 2026 \cite{villalobos2022will}. The growing reliance on combining existing datasets or leveraging model-generated data rather than curating new datasets underscores the increasing scarcity of publicly available data. Given that established scaling laws suggest larger datasets generally yield superior performance, this shortage could soon pose a significant bottleneck to the continued development of LLMs.

\subsection{LLM-Empowered IoT Architecture}

IoT devices continuously generate real-time, private, multimodal data (e.g., images, speech, and sensor data), yet they require powerful AI capabilities to manage resources efficiently and deliver intelligent services to users. On the other hand, as mentioned previously, LLMs possess exceptional generalization and reasoning capabilities but face limitations due to data scarcity. By bridging this gap, a synergistic integration between IoT and LLMs can unlock new possibilities for intelligent applications. 

In alignment with the IoT vision for the 6G era, we propose an LLM-empowered IoT architecture to overcome the aforementioned challenges. As shown in Fig. 1, the architecture integrates LLMs within IoT ecosystems across the space–air network and the ground network to form a collaborative and hierarchical intelligence system. The cloud server, as a centralized intelligence hub, is responsible for network management and LLM updates. Depending on device capability, LLMs are either split and collaboratively deployed between edge servers and IoT devices, or compressed into lightweight versions for direct on-device inference to reduce reliance on the cloud computing for real-time decision-making, automation, and data analytics. 

The synergy of LLM and IoT in the proposed architecture is two-fold:  LLM for 6G IoT and LLM on 6G IoT. LLM for 6G IoT refers to the use of LLMs to enhance the intelligence and functionality of IoT systems in 6G environments, enabling smarter applications and more efficient network operations.LLM on 6G IoT refers to the deployment and execution of LLMs over 6G-enabled IoT infrastructure, focusing on how LLMs can run efficiently on resource-constrained devices with the support of 6G networks.  In the following, we will illustrate the basic ideas of LLM for 6G IoT in Section III and LLM on 6G IoT in Section IV, respectively.

\section{LLM for 6G IoT}
As shown in Fig. 2, LLMs for 6G IoT can be categorized into two key aspects: LLMs-empowered IoT applications and LLMs-enhanced network management.

\subsection{LLM-Empowered IoT Applications}
With the advancement of 6G networks, edge computing, and AI algorithms, the application of LLMs in the IoT field will be more extensive, providing strong support for the future smart society. As shown in Fig. 2, three representative applications are introduced as follows:
\begin{itemize}
	\item \textit{Smart healthcare}: IoT is enhancing healthcare by improving real-time monitoring. LLMs can parse medical records, medical papers, and clinical data to assist doctors in making diagnoses and recommending the best treatment options to improve the accuracy of medical decisions. Combined with data from wearable devices, LLM can analyze a patient's health status in real-time, predict disease risk, and provide personalized health advice for early disease warning.
	\item \textit{Smart home}: LLMs can enhance the interactivity and autonomy of smart homes, making home devices smarter and more personalized. With the powerful NLP capabilities, LLM can be used in smart speakers and voice assistants (e.g., Alexa, Google Assistant) to achieve smoother, context-aware natural language understanding, enabling users to control home appliances, lights, security systems, etc., through voice or text commands. By learning from long-term user data, LLM can predict user habits to enhance the user experience, such as automatically adjusting the temperature and lighting. Combining computer vision and sensor data, LLM can analyze data from home cameras and door lock sensors to identify abnormal behavior and improve home security.
	\item \textit{Smart city}: IoT plays a crucial role in the development of smart cities by enabling intelligent infrastructure, real-time monitoring, and efficient resource management. LLMs can improve the efficiency of urban management and promote the development of smart cities by understanding and analyzing urban big data. For example, based on traffic sensors, cameras, and GPS data from IoT, LLM can analyze urban traffic flow in time, optimize traffic light scheduling, reduce congestion, and improve commuting efficiency.
\end{itemize}

\subsection{LLM-Enhanced Network Management}
IoT empowered by LLMs facilitates self-adaptive optimization through dynamic power adjustment, transmission strategy optimization, and intelligent task scheduling, thereby enhancing system robustness and energy efficiency. The main processes are shown in Fig. 2. Firstly, IoT devices (e.g., smart sensors, cameras, communication terminals) continuously generate multimodal information such as logs, location information, signal data, sensor data, etc., and send them to edge computing nodes for processing. Edge servers pre-process data from IoT devices and extract key information to reduce the overhead and latency of data transmission. For example, the edge server estimates the channel from the IoT device to the edge server based on the information provided by a local LLM, such as location, logs, and signal strength. Due to the dynamic and changing nature of the network, the resource status information received at the cloud server is not real-time and, therefore, needs to be predicted at the edge based on the available data. After the cloud server receives all the global information and the provided prompts (e.g., optimization goals), it uses the powerful reasoning capability of the LLMs to reason and generate resource optimization strategies. The strategies are cascaded down, and each edge server and IoT device adjusts its resource occupancy according to the strategies.

In contrast to most traditional approaches based on optimization theory techniques that only work with appropriate mathematical models, LLMs can be adapted to different scenarios by learning directly from data without explicit mathematical modeling.

\section{LLM on 6G IoT}
This section explores the key technologies that enable LLMs to deploy on IoT.

\subsection{Edge LLM Fine-Tuning}
LLMs pre-trained in the cloud are deployed within IoT environments for fine-tuning, enabling them to learn domain-specific knowledge from IoT-generated data and adapt to real-world applications more effectively.

\subsubsection{Parameter-Efficient Fine-Tuning}
Considering the limited resources in IoT devices and edge servers, the full fine-tuning (i.e., training all parameters) of LLMs raises the concern of insufficient computing resources. Moreover, full fine-tuning is accompanied by high communication costs for transmitting the full model parameters in federated learning. To this end, the parameter-efficient fine-tuning (PEFT) can be implemented. PEFT reduces computational requirements by updating only a portion of the parameters and freezing most of the parameters of the pre-trained model. As shown in Fig. 3, the PEFT can be categorized into three types: Additive PEFT, selective PEFT, and reparameterization PEFT. Specifically, additive PEFT inserts a minimal number of trainable parameters strategically positioned within the model architecture. Selective PEFT selects the subset of the original parameters for training. Reparameterization PEFT reformulates a model’s architecture by transforming its parameters. For example,  low-rank adaptation (LoRA) \cite{hu2021lora} decomposes pre-trained weights into low-rank matrices for updating.
\begin{figure*}[t]
	\centering
	\includegraphics[width=0.85\textwidth]{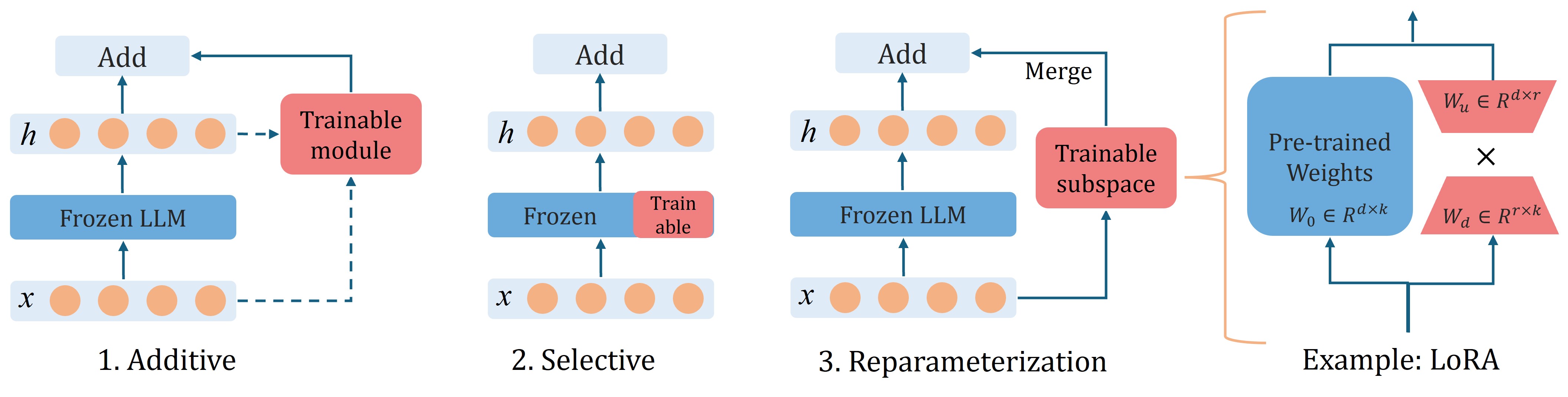}
	\caption{Parameter-efficient fine-tuning}
	\label{PEFT}
\end{figure*}
\begin{figure*}[h]
	\centering
	\includegraphics[width=0.92\textwidth]{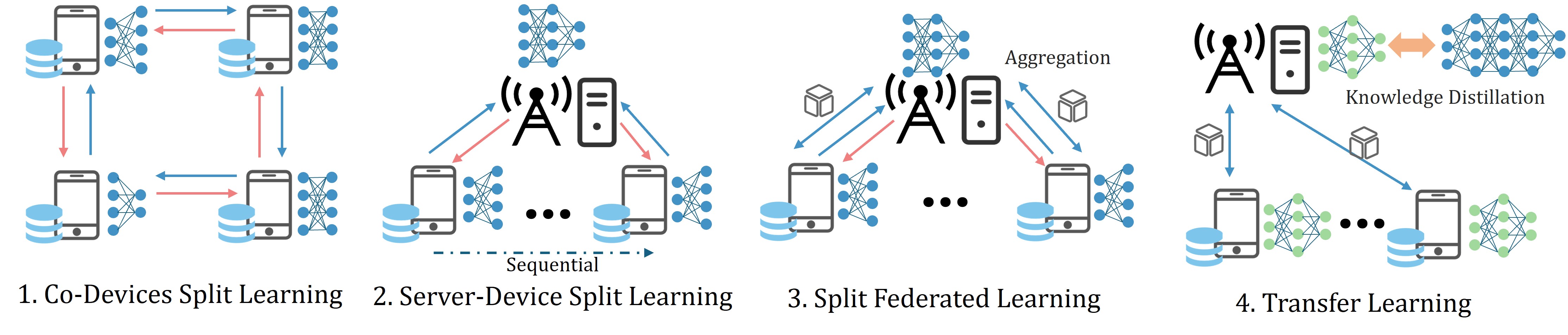}
	\caption{Distributed learning frameworks for edge fine-tuning.}
	\label{DL}
\end{figure*}

Although PEFT methods can reduce computational overhead, they require storing intermediate activations to compute the gradients of the trainable parameters, which need a significant memory requirement (over 70\% full fine-tuning). For most IoT devices, on-device LLM fine-tuning is still resource-intensive. On the other hand, collecting the raw data from IoT devices to an edge server raises privacy concerns.

\subsubsection{Distributed Learning Framework}
Distributed learning with multi-device collaboration or server-device collaboration is a promising solution. Fig. 4 demonstrates the distributed learning frameworks for edge fine-tuning. 
\begin{itemize}
	\item Collaborative devices split learning: The pre-trained model is partitioned into multiple submodels deployed in a set of devices within the trusted domain. Data from all devices is sent to the device with the model's input layer, which is trained through a well-designed pipeline mechanism.
	\item Server-device split learning: The pre-trained model is split into a client-side submodel and a server-side submodel. After completing a client's training, the client uploads the client's model to the server to update the overall model. After that, the server splits the model and sends it to the next client to continue training.
	\item Split federated learning: Split federated learning (SFL) combines the principles of federated learning (FL) and split learning (SL) by splitting the model between the client and server, allowing clients to train part of the model locally while offloading the remaining computation to the server, thus enhancing privacy and reducing client-side memory usage.
	\item Transfer learning: IoT devices are deployed with lightweight LLMs or other small models, and edge servers are deployed with LLMs. Knowledge transfer between LLMs and small models is accomplished through knowledge distillation.
\end{itemize}

A shared objective of the aforementioned distributed learning approaches is to alleviate the memory burden and computational demands on devices. Distributed learning methods combined with PEFT can effectively reduce the memory footprint, computational requirements, and communication overhead of model aggregation.

\subsection{Edge LLM Inference}
Existing cloud-based LLM inference may expose raw data to potential leakage risks, especially during transmission or third-party processing. Moreover, cloud-based inference will introduce significant latency that will defy the need for LLMs to serve IoT. 

\subsubsection{On-Device Inference} 
A feasible solution for fast LLM inference is on-device inference. The model is lightened by quantizing and pruning the large model, and then the knowledge of the original large model is obtained through knowledge distillation. With lightweight LLM, IoT infers at the device to get results. The model compression is accompanied by a decrease in accuracy, which is intolerable for some applications that require high precision.

\subsubsection{Co-Inference} 
Collaborative inference is incorporated into our proposed architecture to ensure accuracy and the resource constraints of IoT devices. Collaborative inference alleviates the computational burden on IoT devices by offloading workloads to a server through layer-wise model partitioning. Beyond ensuring strong privacy preservation, it can also reduce communication overhead when the size of the intermediate features at the partitioned layer is smaller than that of the raw data, making it a more efficient approach for resource-constrained IoT environments. Furthermore, similar to multi-device split learning, a multi-hop collaborative inference architecture can be designed based on the specific network topology and device resources. 

\begin{figure*}
	\centering
	\includegraphics*[width=0.8\textwidth]{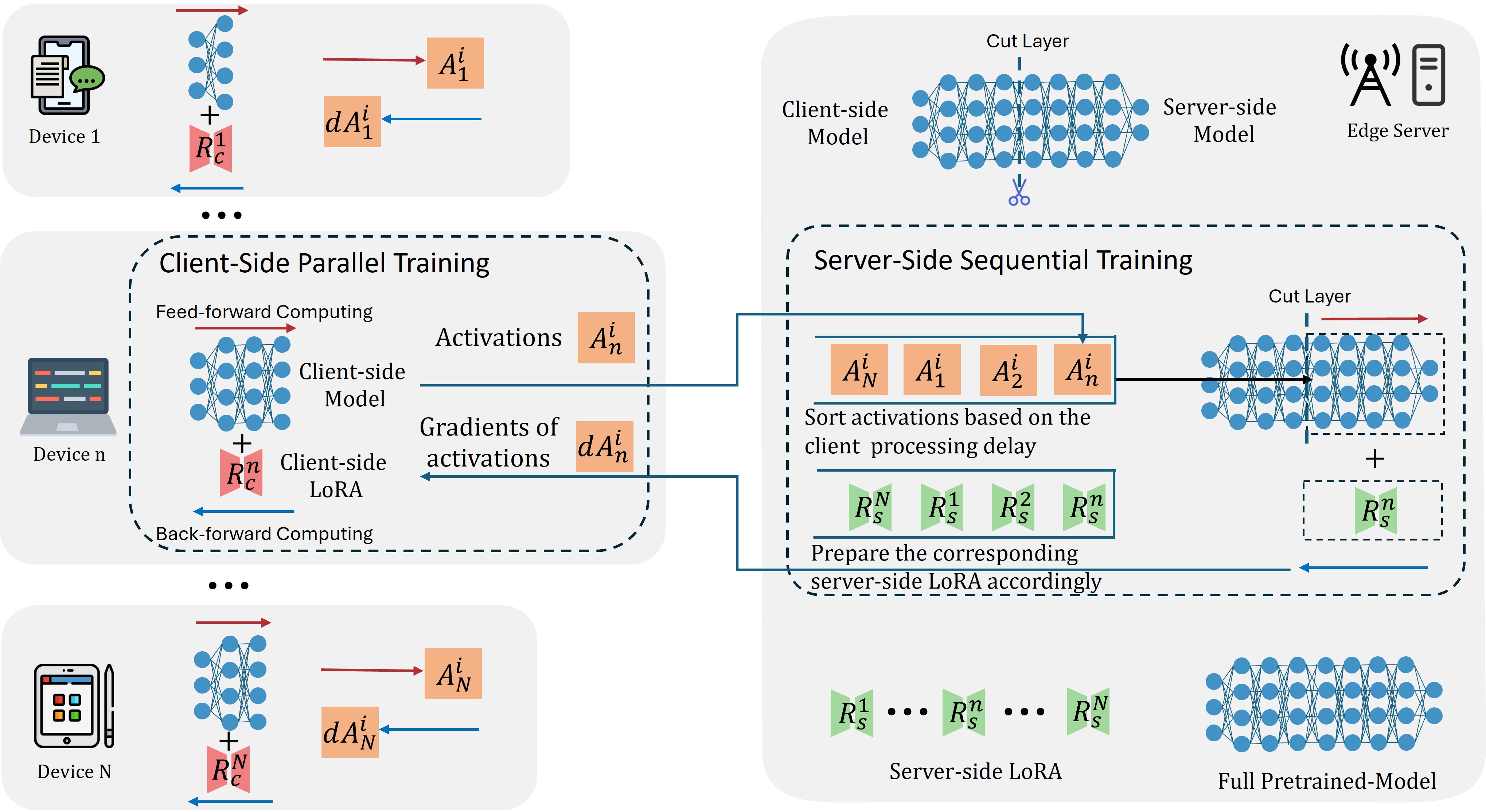}\\
	\caption{An illustration of the memory-efficient SFL framework.}\label{fig.workflow}
\end{figure*}

\section{Memory-Efficient Split Federated Learning for LLM Fine-Tuning} 
We consider a typical two-tier IoT network, which consists of an edge server and a set of IoT devices (clients). The edge server, with more powerful computational and memory capability, is primarily responsible for the model fine-tuning training task and manages model aggregation and split.   These mobile devices are heterogeneous, with different computing capabilities and memory. Each client has a local dataset for fine-tuning and the local datasets of the clients are non-independent and identically distributed (Non-IID). The edge server and the IoT devices collaboratively fine-tune a transformer-based LLM by the LoRA method \cite{hu2021lora} for a specific downstream task.

As shown in Fig.~\ref{fig.workflow}, each client is assigned a client-side sub-model that includes the embedding layer and a limited number of shallow transformer blocks, along with the corresponding LoRA adapters. The server, maintaining a full pre-trained model and multiple server-side LoRA adapters corresponding to the client-side LoRA adapters,  is responsible for executing the forward and backward propagation of those layers. The model is partitioned at the granularity of transformer blocks to flexibly balance the computational and memory load between the client and the server. Additionally, the server manages the synchronization of LoRA adapters by periodically aggregating updates from both the client-side and server-side adapters. The goal is to collaboratively learn an optimized LoRA model that minimizes the global loss function across all participating clients. The training procedure can be divided into client-side parallel training  and server-side sequential training:
\begin{itemize}
	\item  Client-side parallel training:  Each client performs forward propagation and backward propagation of the client-side pre-trained model locally. After the client completes the forward propagation, the activation, the corresponding label, and the corresponding index of the split layer are uploaded to the server via wireless communication. 
	\item Server-side sequential training: On the server, the system implements a processing sequence scheduling module to sort and prioritize incoming smashed data according to each client's processing delay. The corresponding server-side LoRA adapters are loaded, and the received activations are passed into the corresponding cut layer for forward propagation and backward propagation in the remaining model. 
\end{itemize} 

After completing the given rounds of training, clients transmit their locally updated client-side LoRA adapters to the server via wireless communication. These client-side LoRA adapters are paired with their corresponding server-side adapters to form complete LoRA adapters for aggregation. The server then applies the FedAVG algorithm \cite{mcmahan2017communication} to aggregate all clients’ full LoRA adapters into a unified global adapter. After aggregation, the server partitions the global LoRA adapters to extract the client-side components, which are subsequently distributed back to each client to update their local LoRA adapters.

In the proposed framework, client-side submodels are trained in parallel, while server-side submodels are trained sequentially. The order of sequential computation impacts the backward propagation of the clients. The decision of client order for server-side training influences the overall training time. Therefore, to minimize the overall completion time, the key is to hide the communication time and the client computation time under the server computation time as much as possible \cite{9155272}. Since the gradient size of each layer is the same and the gradient transmission time is much smaller than the backward propagation time. Therefore, we use a greedy algorithm where the server prioritizes tasks with a longer backward propagation time of the client.

\section{Case Study} 
This section presents a case study to demonstrate the effectiveness of the proposed memory-efficient SFL framework.

\subsection{Considered Scenario}

We use an RTX 4080s server with a computational capability of 52.2 TFLOPS and consider six heterogeneous clients: a Jetson Nano (0.472 TFLOPS) with first one transformer layer, a Jetson TX2 (1.33 TFLOPS) with first one transformer layer, a Snapdragon 8s Gen 3 (1.689 TFLOPS) with first two transformer layers, a Snapdragon 8 Gen 3 (2.774 TFLOPS) with first two transformer layers, an A17 Pro (2.147 TFLOPS) with first three transformer layers, and an M3 (3.533 TFLOPS) with first three transformer layers. The data rate between the server and each client is set to 100 Mbps. We leverage BERT-base \cite{devlin2018bert} as the pre-trained model for text analysis tasks using CARER dataset \cite{saravia-etal-2018-carer}. We set the LoRA rank, batch size, learning rate,  maximum sequence length, and target accuracy to 16, 16, 0.00001, 128, and 0.89. We compare the proposed scheme with the following baselines: 1) SL \cite{10040976}; 2)SFL \cite{tian2022fedbert}.

\begin{figure}
	\centering
	\includegraphics[width=0.45\textwidth]{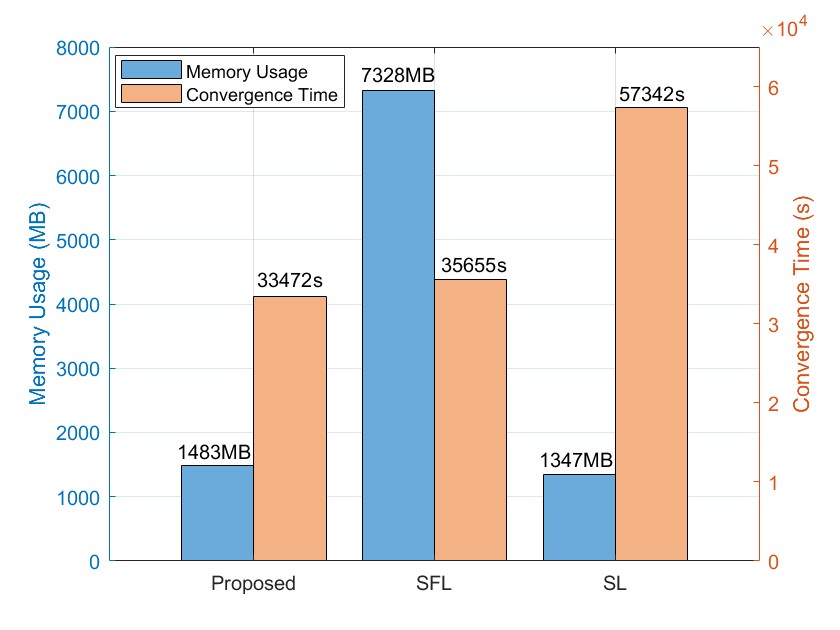} 
	\caption{Comparison of memory usage and convergence time across different schemes.}  
	\label{fig:subfig1}
\end{figure}

\subsection{Simulation Results}

Figure 6 shows the memory usage and the convergence time of the three frameworks. Although SL has the smallest memory usage, it results in the longest convergence time. SFL takes up a very large amount of memory due to the need to maintain multiple large models. The proposed scheme is able to achieve the fastest convergence. Compared with SL, the proposed scheme reduces the training time by 40\% at the 10\% memory cost. Compared to SFL, the proposed scheme reduces 79\% of memory and 6\% of training time.   The proposed scheme reduces the memory footprint by reusing a full LLM for sequential training and reduces the overall training time by designing a reasonable training scheduling scheme.

\section{Open Issues}

\subsection{Resource-Aware LLM Deployment} 
IoT devices with limited memory, computational power, and processing capabilities, typically operate in energy-constrained environments, which significantly impacts the feasibility of running LLMs. The high memory consumption and computational resources of LLMs exceed the storage capabilities of many IoT devices, preventing them from loading full models locally. Given the sheer number of connected IoT devices, bandwidth constraints further restrict efficient communication and data exchange. Moreover, IoT networks consist of devices with vastly different resources, leading to imbalanced workloads. 

To address the limited and heterogeneous resources of IoT devices, a combination of model optimization techniques, distributed computing strategies, and efficient workload management is essential. Model compression methods such as quantization, pruning, and knowledge distillation can significantly reduce the resource requirements of LLMs. Additionally, split computing allows IoT devices to offload heavy computations to edge servers, enabling a balance between local processing and remote inference. Furthermore, adaptive workload distribution ensures that high-performance edge nodes handle complex tasks while lightweight devices process simpler workloads, optimizing overall system efficiency.

\subsection{On-Demand LLM Deployment}
Deploying LLMs in IoT environments is complicated by constrained resources and varying task demands. Centralized cloud-based deployment introduces high latency and bandwidth costs, which are unsuitable for delay-sensitive applications like real-time monitoring and smart healthcare. Moreover, in IoT applications, service requests of LLMs are diverse and random, and the computational demands depend on the complexity of the task, making it inefficient to allocate fixed resources for model execution continuously. 

To enable the on-demand deployment of LLMs, a combination of edge caching and model compression is essential. Edge caching enables frequently used model components or inference results to be stored closer to IoT devices, reducing redundant computations and minimizing communication overhead. Meanwhile, a trade-off between model compression and performance must be carefully managed \cite{yang2024adaptive}, as highly compressed models may introduce accuracy degradation, while full-scale models demand significant computational resources.

\subsection{Data Privacy and Security}
IoT devices collect sensitive data (such as personal, health, and location information) raising serious privacy and security concerns, especially when relying on cloud-based processing. Many devices lack secure storage and are exposed to risks like data breaches, encryption vulnerabilities, and data poisoning, particularly in heterogeneous networks with inconsistent security standards. Furthermore, decentralized AI models are susceptible to attacks such as model inversion, threatening data integrity, and regulatory compliance \cite{chen2024unveiling}.

To mitigate privacy and data security risks, privacy-preserving strategies are essential in collaborative learning systems. Differential privacy provides formal guarantees by injecting calibrated noise into shared information, thereby reducing the risk of input reconstruction. Additionally, U-shaped split learning enhances data confidentiality by relocating both the input and output layers to the client side, effectively preventing label leakage. These techniques collectively offer complementary safeguards that strengthen data protection throughout the AI training and inference lifecycle.

\section{Conclusion}
In this article, we have proposed an LLM-empowered IoT architecture to support IoT applications and facilitate intelligent network management. LLM for 6G IoT is to enhance device intelligence and efficient resource management in dynamic IoT environments. LLM on 6G IoT focuses on optimizing infrastructure to support LLM deployment, leveraging edge computing and efficient networking to accommodate the high computational and communication demands of LLMs. In addition, we proposed a memory-efficient SFL framework for LLM fine-tuning to reduce memory footprint and training time. Looking ahead, we’ve outlined open issues that need to be addressed to fully realize the promise of LLM-empowered 6G IoT. With continued innovation, the convergence of 6G, IoT, and LLMs could become the foundation for intelligent future networks.

\section*{Acknowledgments}
This work was supported in part by the Pengcheng Laboratory Major Key Project under Grants 2025QYA002, PCL2023AS1-5, and PCL2024A01, in part by the Natural Science Foundation of China under Grants 62201311, 62201071 and 62192712, in part by the Young Elite Scientists Sponsorship Program by CAST under Grant 2023QNRC001.

\normalem
\footnotesize
\bibliographystyle{IEEEtran}
\bibliography{IEEEabrv,references}

 \section*{Biographies}
 \textbf{Xiaopei Chen} [S'25] (ftchenxp@mail.scut.edu.cn) received the B.S. degree in electronics and information engineering and the M.S. degree in electronics and communication engineering from Fuzhou University, Fuzhou, China, in 2020 and 2023, respectively. He is currently a joint Ph.D. student in Information and Communication Engineering with the School of Future Technology, South China University of Technology, Guangzhou, China, and the Pengcheng Laboratory, Shenzhen, China.  His current research interests include edge intelligence and vehicular networks.

\textbf{Wen Wu} [S'13-M'20-SM'22] (wuw02@pcl.ac.cn) earned his Ph.D. degree in electrical and computer engineering from the University of Waterloo, Ontario, Canada, in 2019. He received his B.E. degree in information engineering from South China University of Technology, Guangzhou, China,and his M.E. degree in electrical engineering from University of Science and Technology of China, Hefei, in 2012 and 2015, respectively. He worked as a postdoctoral fellow with the Department of Electrical and Computer Engineering, University of Waterloo. He is currently an associate researcher at Pengcheng Laboratory, Shenzhen, China. His research interests include 6G networks, pervasive network intelligence, and network virtualization.

\textbf{Liang Li} [S’19-M’21] (lil03@pcl.ac.cn) received the Ph.D. degree in the School of Telecommunications Engineering at Xidian University, China, in 2021. She was a post-doctoral faculty member with the School of Computer Science (National Pilot Software Engineering School), Beijing University of Posts and Telecommunications, from 2021 to 2023. Since 2023, she has been with Pengcheng Laboratory, China, where she is currently an assistant researcher with the Department of Advanced Interdisciplinary Research. She was also a visiting Ph.D. student with the Department of Electrical and Computer Engineering, University of Houston, Houston, TX, USA, from 2018 to 2020. Her research interests include edge intelligence, federated learning, edge computing and caching, data-driven robust optimization, and differential privacy.

 \textbf{Fei Ji} [M'06] (eefeiji@scut.edu.cn) received the B.S. degree in applied electronic technologies from Northwestern Polytechnical University, Xi’an, China, in 1992, and the M.S. degree in bioelectronics and the Ph.D. degree in circuits and systems from the South China University of Technology, Guangzhou, China, in 1995 and 1998, respectively. She was a Visiting Scholar with the University of Waterloo, Waterloo, ON, Canada, from June 2009 to June 2010. She worked with the City University of Hong Kong, Hong Kong, as a Research Assistant from March 2001 to July 2002 and a Senior Research Associate from January 2005 to March 2005. She is currently a Professor with the School of Electronic and Information Engineering, South China University of Technology. Her research focuses on wireless communication systems and networking.

\end{document}